\documentstyle[12pt]{article}
\newcommand{\rf}[1]{(\ref{#1})}
\newcommand{\beq}{\begin{equation}}
\newcommand{\eeq}{\end{equation}}

\newcommand{\bea}{\begin{eqnarray}}
\newcommand{\eea}{\end{eqnarray}}

\newcommand{\bC}{{\bf C}}

\newcommand{\cT}{{\cal T}}

\newcommand{\cH}{{\cal H}}
\newcommand{\cV}{{\cal V}}

\newcommand{\pa}{\partial}

\newcommand{\noi}{\noindent}

\newcommand{\bR}{{\bf R}}

\begin{document}
\topmargin 0pt
\oddsidemargin 5mm
\headheight 0pt
\topskip 0mm

\addtolength{\baselineskip}{0.20\baselineskip}

\pagestyle{empty}

\hfill August 1993

\begin{center}

\vspace{12pt}
{\Large \bf
Classification and construction of unitary topological field theories
in two dimensions\footnote{Supported in part by a NATO Science
collaboration grant.}}



\vspace{2 truecm}

{\em Bergfinnur Durhuus\footnote{e-mail: durhuus@math.ku.dk}}

\medskip

Matematisk Institut, University of Copenhagen \\
Universitetsparken 5, 2100 Copenhagen \O \\
Denmark

\vspace{1.3 truecm}

{\em Thordur Jonsson\footnote{e-mail: thjons@raunvis.hi.is}}

\medskip

Raunvisindastofnun Haskolans, University of Iceland \\
Dunhaga 3, 107 Reykjavik \\
Iceland

\vspace{2 truecm}

\end{center}

\noi
{\bf Abstract.} We prove that unitary two-dimensional
topological field theories are uniquely characterized by $n$
positive real numbers
$\lambda _1,\ldots \lambda _n$ which can be regarded as the
eigenvalues of a hermitean handle creation operator.  The number $n$ is
the dimension of the Hilbert space associated with the circle and the
partition functions for closed surfaces
have the form
$$
Z_g=\sum_{i=1}^{n}\lambda _i^{g-1}
$$
where $g$ is the genus.
The
eigenvalues can be arbitary positive numbers.
We show how such a theory can be constructed
on triangulated surfaces.

\vfill

\newpage
\pagestyle{plain}

\section{Introduction}
Topological quantum field theory (TQFT) \cite{w1,at}
has given
considerable insight into a number of unsolved
problems in theoretical
physics \cite{w3,w5}
and also provided an important research tool in pure mathematics
\cite{w2,rt,tv}.
Most work in this field has been devoted to studying particular examples
of such theories and uncovering their physical content as well as their
relation to low-dimensional topology.

The problem of classifying topological quantum field theories
is a hard one in more than two dimensions since
it is intimately tied up with the classification of
topological manifolds in these dimensions.  In two dimensions the
situation is radically different since a compact
orientable two-dimensional manifold is
characterized topologically
by its genus and the number of boundary components.  In \cite{w5} it was
shown that the two-dimensional theories have indeed a very simple structure.

The
classification problem for TQFT in two dimensions was addressed in
\cite{th1}, see also \cite{th2},
using triangulated surfaces and fields associated with
vertices.  In these papers topological
theories were constructed with the partition
function for a closed surface of genus $g$ given by
\beq
Z_g=\lambda ^{g-1}\label{partition}
\eeq
where $\lambda$ is an arbitrary
positive number.

In the present paper we prove
that the most general unitary
TQFT in two dimensions is in fact a direct sum of
theories of this type.
The proof is very simple.  It  uses the fact that
any two-dimensional surface can be constructed by gluing together
spheres with three or fewer boundary components and therefore
any unitary TQFT in two dimensions
is determined by the partition functions for a sphere
with three or fewer holes.  Unitarity implies that an associated
handle creation operator is hermitean and it follows that the
theory
is characterized up to equivalence by the values of partition functions
for closed surfaces.

The classification of unitary TQFT on triangulated surfaces with the fields
defined on links and taking values in a finite set
has been discussed in a number of recent papers
\cite{bachas,filk,fukuma}.   It was first shown in \cite{bachas} that the
weight factors for triangles can in this case
be regarded as structure constants
of a semisimple associative algebra and the physical Hilbert space can
be identified with the center of the algebra.  It follows that in these
theories the number $\lambda$ in \rf{partition} is always
the inverse square of
an integer.  This is also the case in two-dimensional topological gauge
theories \cite{jwtop,wtop}.
Theories with fields defined on links can easily be realized as
vertex theories.

In the next section we describe the Atiyah axioms for unitary
TQFT and point out
simplifications that occur in two dimensions.  We then proceed to
show that the theory is determined by the three loop function on the
sphere and use the gluing axiom
to show that the partition function for a closed surface is necessarily
a sum of exponentials as in \rf{partition}.  In section 4
we construct a lattice TQFT of the most general form. In the
final section we comment on the classification problem in higher
dimensions.

\section{The Atiyah axioms}
In this section we recall the axioms for a unitary TQFT \cite{at} as
they apply in two dimensions.  We assume that all manifolds are smooth,
oriented and
compact
unless otherwise is stated.  If $M$ is an oriented manifold,
we denote by $M^*$ the same manifold with the orientation reversed.

A unitary TQFT in two dimensions
comprises the following assignments: To each closed
1-dimensional manifold
$\Sigma$ there is assigned a finite dimensional Hilbert space $\cH
_{\Sigma}$ with inner
product $<,>_{\Sigma}$.  To each surface $S$ (not necessarily closed)
there is assigned an element $Z(S)\in \cH _{\Sigma}$, called
the partition function for $S$, where $\Sigma =\pa
S$ is the boundary of $S$.  The Hilbert space associated with the empty
1-dimensional manifold $\emptyset$ is $\cH _{\emptyset}= \bC$ so $Z(S)\in
\bC$ if $S$ is closed.
These objects satisfy the following conditions:
\begin{enumerate}
\item If $f:\Sigma _1\mapsto \Sigma _2$ is an orientation preserving
diffeomorphism between 1-manifolds then there is an associated unitary mapping
\beq
U_f:\cH _{\Sigma
_1} \mapsto \cH _{\Sigma _2}
\eeq
 such that $U_{g\circ f}=U_gU_f$ if $g$ is
an orientation preserving diffeomorphism from $\Sigma _2$ to another
1-manifold $\Sigma
_3$.

If $f$ extends to an orientation preserving diffeomorphism between
surfaces, $S_1\mapsto
S_2$, where $\pa S_i=\Sigma _i$, $i=1,2$, then $U_f(Z(S_1))=Z(S_2)$.

\item For any 1-manifold $\Sigma$,
\beq
\cH _{\Sigma ^*}=\cH _{\Sigma}^*
\eeq
i.e. we have an identification between these spaces, which
is equivalent to giving a
non-degenerate bilinear form
\beq
( , )_{\Sigma}:\cH _{\Sigma}\times \cH _{\Sigma ^*}\mapsto \bC
\label{bil}
\eeq
such that
\beq
(x,y)_{\Sigma}=(y,x)_{\Sigma ^*}.\label{a}
\eeq
This form is preserved by diffeomorphisms, i.e.
for $f:\Sigma _1\mapsto\Sigma _2$ as in axiom 1,
\beq
(x,y)_{\Sigma _1}=(U_fx,U_{f^*}y)_{\Sigma _2} \label{b}
\eeq
for all $x\in\cH _{\Sigma _1}$, $y\in \cH _{\Sigma _1^*}$, where
$f^*:\mapsto\Sigma _1^*\mapsto \Sigma _2^*$ denotes $f$
regarded as an orientation preserving map between $\Sigma _1^*$
and $\Sigma _2^*$.

Defining the conjugate linear isomorphism $y\mapsto y^*$ from $\cH _{\Sigma}$
to $\cH _{\Sigma ^*}$
 by
\beq
(x,y^*)_{\Sigma}=<x,y>_{\Sigma}  \label{c}
\eeq
for $x,y\in \cH _{\Sigma}$, we furthermore assume that
$x^{**}=x$ for all $x\in \cH _{\Sigma}$ and
\beq
Z(S^*)=Z(S)^*  \label{con}
\eeq
for any surface $S$ with boundary $\Sigma$.

\item  If $\Sigma =\Sigma _1\cup \Sigma _2$ is a disjoint union of
1-manifolds then
$\cH _{\Sigma}=\cH _{\Sigma_1}\otimes \cH _{\Sigma_2}$ and
we have a corresponding factorization of the bilinear forms
and unitary maps, i.e. $(,)_{\Sigma}=(,)_{\Sigma _1}
\otimes
(, )_{\Sigma _2}$, and if $f_1:\Sigma _1\mapsto
\Sigma _1'$ and $f_2:\Sigma _2\mapsto \Sigma _2'$ are orientation
preserving diffeomorphisms, then $U_f=U_{f_1}\otimes U_{f_2}$ where $f:
\Sigma \mapsto \Sigma _1'\cup \Sigma _2'$ denotes the diffeomorphism that
equals $f_1$ on $\Sigma _1$ and $f_2$ on $\Sigma _2$.

Let $S_1$ and $S_2$ be two surfaces such that $\pa S_1=\Sigma
_1\cup\Sigma _3$ and $\pa S_2=\Sigma _2\cup \Sigma _3^*$.  Let $S$ be the
surface obtained by gluing $S_1$ and $S_2$ together along their $\Sigma
_3$ boundary component,
\beq
S=S_1\cup _{\Sigma _3}S_2.
\eeq
Then $Z(S)=(Z(S_1),Z(S_2))_{\Sigma _3}$ where, by abuse of notation,
 $(\cdot ,\cdot )_{\Sigma _3}$
denotes the  pairing
\beq
\cH _{\Sigma _1}\otimes \cH _{\Sigma _3}\otimes \cH _{\Sigma _2}
\otimes \cH _{\Sigma
_3}^*\mapsto \cH _{\Sigma _1}\otimes \cH _{\Sigma _2}
\eeq
induced by \rf{bil}.

Concretely, if $\{ x_i\}$, $\{ y_j\}$, $\{ z_k\}$ are bases for $\cH
_{\Sigma _1}$,  $\cH
_{\Sigma _2}$, $\cH
_{\Sigma _3}$, respectively, and $\{ z_k^*\}$ is the dual basis for $\cH
_{\Sigma _3}^*$, then we can write
\bea
Z(S_1)&=&\sum _{i,k}c_{ik}\, x_i\otimes z_k\\
Z(S_2)&=&\sum _{j,l}d_{jl}\, y_j\otimes z_l^*
\eea
for suitable constants $c_{ik}$, $d_{jl}$, and
\beq
Z(S)=\sum _{i,j,k}c_{ik}d_{jk}\, x_i\otimes y_j.
\eeq

\item Let $\Sigma$ be an oriented 1-manifold and orient the cylinder
$\Sigma\times [ 0,1] $ such that $x\mapsto (x,0)$ is an
orientation reversing map from $\Sigma$ onto $\Sigma\times \{ 0\}$
whereas $x\mapsto (x,1)$ is an
orientation preserving mapping from $\Sigma$ onto $\Sigma\times \{ 1\}$.
Using the canonical identification
$\cH _1^*\otimes \cH _2\approx {\rm Hom} (\cH _1,\cH _2)$,  we may according to
axioms 2 and 3 regard $Z(\Sigma\times [ 0,1] )$ as a linear map from
$\cH _{\Sigma\times \{ 0\} }$ to $\cH _{\Sigma\times \{ 1\} }$ and we
assume that
$
Z(\Sigma\times [ 0,1] )=U_f
$
where $f:\Sigma\times \{ 0\}\mapsto \Sigma\times \{ 1\} $ is defined
by $f((x,0))=(x,1)$.
Using the mapping $U_f$
to identify the spaces $\cH _{\Sigma \times\{ 0\} }$ and
$\cH _{\Sigma\times \{ 1\} }$ we can
write
\beq
 Z(\Sigma\times [ 0,1])  =I
\eeq
where $I$ is the identity mappping.

\end{enumerate}

We refer to \cite{at} for a detailed discussion of the axioms.
Here we make a few comments that apply especially in the two-dimensional
case.

It is a standard consequence of the axioms that the unitary mapping
$U_f$ with $f$ as in axiom 1 only depends on the homotopy class of the
diffeomorphism $f$.  If $\Sigma _1$ and $\Sigma _2$ are connected, i.e.
circles, there is
only one homotopy class of orientation preserving diffeomorphisms and we can
write $U_f=U(\Sigma _1,\Sigma _2)$.  By axiom 1 it follows that
the mappings $U(\Sigma _1,\Sigma _2)$ yield a canonical identification of
all the Hilbert spaces $\cH _{\Sigma}$ (where $\Sigma$ is
connected) with a single
Hilbert space which we denote by $\cH$ with inner product $<,>$.  With these
identifications all the mappings $U(\Sigma _1,\Sigma _2)$
are of course the identity and by
\rf{a} and \rf{b} a unique symmetric bilinear form $(,)$ is defined on $\cH$.
Similarly, by \rf{b} and \rf{c} the $*$-maps define a unique conjugate
linear involution $x\mapsto x^*$ on $\cH$ given by
\beq
(x,y^*)=<x,y>. \label{nbil}
\eeq

It follows from the assumed factorization properties of the inner products,
bilinear forms and unitary mappings that the vectorspace associated to a
1-manifold with $n$ boundary components $\Sigma _1,\ldots ,\Sigma _n$ can
be identified with $\cH ^{\otimes n}$.  If $f$ is a diffeomorphism
of $\Sigma _1\cup \ldots \cup \Sigma _n$ onto itself
which permutes the boundary components,
then the mapping induced by $U_f$ acts on $\cH ^{\otimes n}$ by the
corresponding permutation of factors in the tensor product.  This implies that
if $S$ is a {\em connected} surface with $\pa S=\Sigma _1 \cup \ldots\cup
\Sigma _n$ then $Z(S)\in\cH ^{\otimes n}$ is a symmetric tensor since there
exist orientation preserving
diffeomorphisms that permute the boundary components
of $S$ in any prescribed way.  Moreover, since any surface $S$ possesses an
orientation resversing diffeomorphism, we conclude that $Z(S^*) =Z(S)$ and
hence, by \rf{con}, that
\beq
Z(S)=Z(S)^*   \label{real}
\eeq
for any surface $S$.

Letting $\cH _R$ denote the real subspace of $\cH$ defined by
\beq
\cH _R=\{ x\in\cH : x=x^*\}
\eeq
we have a direct sum decomposition over $\bR$,
\beq
\cH =\cH _R\oplus i\cH _R.   \label{bas}
\eeq
According to \rf{nbil} and the symmetry of $(,)$ it follows that the
restriction of the inner product to $\cH _R$ is a real inner product
on $\cH _R$ which equals the restriction of the bilinear form to $\cH _R$.
We thus conclude from \rf{real} that any two-dimensional unitary TQFT
is effectively real, i.e. we might have started from the outset with
real Hilbert spaces and partition functions satisfying the analogues
of axioms 1-4 without \rf{c} and \rf{con}.

\section{Classification} Let us assume that we are given a unitary
TQFT satisfying the axioms of the previous section.  Let us denote the
partition function for a connected surface of genus $g$ with $n$ boundary
components by $Z_{g,n}$ and write $Z_{g,0}=Z_g$.
We choose an orthonormal basis $\{ x_i\}$ for $\cH _R$ which, according to
\rf{bas}, also constitutes an orthonormal basis for $\cH$ with $x_i^*=x_i$.
 The one loop function on the sphere can
then be expressed as
\beq
Z_{0,1}=\sum_{i}d_i\,x_i.
\eeq
The two and three loop functions on the sphere can similarly be written as
\beq
Z_{0,2}=\sum_{ij}q_{ij}\,x_i\otimes x_j\label{q}
\eeq
and
\beq
Z_{0,3}=\sum _{ijk} C _{ijk}\, x_i\otimes x_j\otimes x_k,
\eeq
where $q_{ij}$ and $C_{ijk}$ are real and symmetric under interchange of the
indices.
By axioms 3 and 4 we have
\beq
\sum _iq_{ij}\,x_i=x_j^*
\eeq
so $q_{ij}=\delta _{ij}$ with respect to the chosen basis.

We define a handle operator
$H\in {\rm End }(\cH )$ by gluing together two three loop functions:
\beq
H=\sum _{il}H_{il}\, x_i\otimes x_l^*
\eeq
where
\beq
H_{il}=\sum _{jk}C_{ijk}C_{ljk}.
\eeq
This is the same as regarding $Z_{1,2}$ as an operator on $\cH$.
Applying $H$ to the vector $Z_{g,1}$ clearly gives $Z_{g+1,1}$.  The
operator $H$ is hermitean by \rf{con} i.e. symmetric on $\cH _R$.

We now choose the basis
$\{ x_i\}$ to consist of the eigenvectors of $H$ and let $\lambda _1, \ldots
\lambda _n$ denote the eigenvalues of $H$ (not necessarily distinct),
$n={\rm dim}\,\cH$.  Then by axiom 3
\bea
Z_g&=&<Z_{0,1},H^{g}Z_{0,1}>\\
   &=&\sum _{i}\lambda _i^{g} |d_i|^2 \label{spec}
\eea
for any $g\geq 0$.
Furthermore,
\beq
Z_{g+1}={\rm Tr} H^g=\sum _i\lambda _i^g\label{spec2}
\eeq
for all $g\geq 0$.
It follows that
\beq
\sum _i\lambda _i^g|d_i|^2=\sum _i\lambda _i^{g-1} \label{spec3}
\eeq
for all $g\geq 1$.

Eq. \rf{spec3} implies that
all the eigenvalues are nonegative.  In order
to prove this we order the eigenvalues according to their absolute value so
that $|\lambda _i|\geq |\lambda _j|$ if $i\leq j$ and assume $\lambda _1\neq
0$.
  Clearly
\rf{spec3} cannot hold for
all $g$ unless $\lambda _1$ is positive and
we conclude also that
\beq
\sum _{i=1}^{k}\lambda _i^g|d_i|^2=k\lambda _1^{g-1}  \label{sub}
\eeq
where $k$ is the multiplicity of $\lambda _1$.  Subtracting \rf{sub}
from \rf{spec3} and repeating the above argument we conclude that all
the eigenvalues are nonegative.
Below we show that a zero eigenvalue cannot occur and
we can choose a basis such that $d_i=\lambda _i^{-{1\over 2}}$.

Consider the operators $C_i$ on $\cH_R$
defined by
\beq
C_i=\sum_{jk}C_{ijk}\,x_j\otimes x_k^*.
\eeq
By the symmetry of $C_{ijk}$ and the four loop
function these operators are mutually commuting
and symmetric so they can be simultaneously diagonalized
and we can choose a new self-dual basis such that
$C_{ijk}=\delta _{ij}\delta _{ik}C_{iii}$.
Using the definition of the handle operator we now find that
$
C_{iii}^2=\lambda _i
$
and $H$ is diagonal in this basis.
If $\lambda _i=0$, then $C_{iii}=0$.
This contradicts axiom 4 which states that
\beq
\sum _{ijk}C_{ijk}d_k\,x_i\otimes x_j^*=I
\eeq
and implies that all $C_{iii}\neq 0$ and $C_{iii}d_i=1$.

It is not hard
to convince oneself that any positive operator can arise as a handle
operator.  In fact we give an explicit construction in
the next section.

We conclude this section by showing that all the information
in a two-dimensional
unitary TQFT is contained in the
spectrum of $H$.  We begin by making precise what we mean by
the equivalence of two unitary TQFT in two dimensions.
Let $T$ be a theory satisfying the axioms of section 2
and let $T'$ be another one whose objects are distinguished from those of
$T$ by a prime.  We say that $T$ and $T'$ are equivalent if for
any 1-manifold $\Sigma$ there exists a unitary mapping
\beq
V_{\Sigma}:\cH _{\Sigma}\mapsto \cH _{\Sigma}'
\eeq
such that the following  conditions hold:
\begin{enumerate}
\item For any orientation preserving diffeomorphism
$f:\Sigma _1\mapsto\Sigma _2$ between 1-manifolds
\beq
U_f'=V_{\Sigma _2}U_fV_{\Sigma _1}^*.
\eeq
\item For any oriented surface $S$
\beq
Z'(S)=V_{\pa S}(Z(S))
\eeq
with the understanding that $Z'(S)=Z(S)$ if $S$ is closed.

\item For any 1-manifold $\Sigma$ we have
\beq
(V_{\Sigma}x,V_{\Sigma ^*}y)_{\Sigma}'=(x,y)_{\Sigma}.
\eeq
\end{enumerate}

Let $T,T'$ be a pair of unitary TQFT's whose handle operators have
the spectrum $\lambda _1, \ldots ,\lambda _n$.
Then, by previous arguments, they
are both equivalent to theories where
all Hilbert spaces for connected boundary components coincide and equal $\cH$
and $\cH '$, respectively.  We have seen that by a
suitable choice of bases the three loop functions then take the form
\beq
Z_{3,0}=\sum _i\sqrt{\lambda _i}\,x_i\otimes x_i\otimes x_i.
\eeq
Let us call such a basis a canonical basis.
Furthermore, $q_{ij}=\delta _{ij}$, $d_i=\lambda _i^{-{1\over 2}}$
with respect to
a canonical basis.  An equivalence between $T$ and
$T'$ is now obtained by mapping a canonical basis for $\cH$ to a canonical
basis for $\cH '$.

We finally remark that two unitary TQFT's whose partition functions
take the same value for all closed surfaces have handle operators
with identical spectra, in view of
\rf{spec2}, and are therefore equivalent.

\section{Construction}
In \cite{th1} a field theoretical construction was given of a TQFT on
triangulated surfaces, using fields defined on vertices, such that
$Z_g=N^{\chi}$, where $N$ could take any positive value and $\chi$ is the
Euler characteristic.
In order to obtain a theory whose partition functions for closed surfaces
are sums of
exponentials we can take a direct sum of theories of this type.

Let us explain what we mean by the
direct sum of two TQFT's.  Suppose we have two theories with loop functions
$Z_{g,n}$ and $Z_{g,n}'$ and Hilbert spaces $\cH$ and $\cH '$ for each
connected boundary component.
The direct sum is a TQFT with the Hilbert space $\tilde{\cH}=\cH\oplus
\cH '$ for each boundary component and loop functions $\tilde{Z}_{g,n}=
Z_{g,n}+Z_{g,n}'$ where we regard $Z_{g,n}+Z_{g,n}'$ in a natural
way as an element of $\tilde{\cH}^{\otimes n}$, $n\neq 0$.
Unitary maps and bilinear forms are defined in the obvious way.
One can then
check that the direct sum satisfies the axioms if the original theories do so.

Here we give an alternative and more direct construction of a TQFT with
an arbitrary handle operator using triangulations and local weights.
Let $S$ be a triangulated surface
of genus $g$ and $N_1,\ldots ,N_n$ a sequence of $n$ not necessarily
distinct positive numbers.  Let $J=\{ 1, \ldots ,n\} $.  A colouring of $S$ is
a
mapping from the vertices of $S$, $\cV (S)$, into $J$.  The image of a vertex
under a colouring is called
its colour.  Given a colouring $\phi$ of $S$ we define
the weight of the vertex $v$ to be $N_i$ where $i=\phi (v)$.  The weight
$w_{\Delta}$ of a triangle $\Delta$ in $S$ with corners whose colours are
$i,j,k$ is defined to be $N_i^{-{1\over 2}}$
if $i=j=k$ but zero otherwise.  The partition
function for a closed surface $S$ is now defined to be
\beq
Z(S)=\sum _{\phi}\prod _{v\in \cV (S)} N_{\phi (v)}\prod _{\Delta \in
\cT (S)}w_{\Delta},
\eeq
where the sum is over all colourings of $S$ and $\cT (S)$ denotes the set of
all triangles in $S$.
It is easy to see that if $S$ is connected the only colourings which contribute
are those that assign the same colour to all vertices and
\beq
Z(S)=\sum _{i=1}^n N_i^{\chi}.
\eeq

Now let $S$ be a connected
triangulated surface with $b$ boundary components.
Let $|\pa S|$ be the number of verticies in $\pa S$ and define
\beq
\zeta _i(S)=N_i^{-{1\over 2}|\pa S|}
\sum _{\phi}\prod _{v\in \cV (S)} N_{\phi (v)}\prod _{\Delta \in
\cT (S)}w_{\Delta},
\eeq
where $\phi$ now runs over all colourings which are $i$ on the boundary.
Clearly $ \zeta _i(S)= N_i^{\chi (S)}$.
We define an $n$-dimensional
Hilbert space $\cH$ by assigning a vector $x_i$ to each colour
and let $\{x_i\}$ be an orthonormal basis for $\cH$ and
we set $x_i^*=x_i$.  The partition function for $S$ is now defined by
regarding $ \zeta _i(S)$ as a coordinate of $Z(S)$ with respect to
the chosen basis, i.e.
\beq
Z(S)=\sum _i \zeta _i(S)x_i^{\otimes b}.
\eeq
It is not hard to check that the theory so defined satisfies all the axioms
and $N_i^{-{1\over 2}}$ are the eigenvalues of the handle operator.

\section{Discussion}
In this paper we have classified all two-dimensional unitary TQFT's
and shown how they can be obtained using locally defined weights on
triangulated surfaces.  Obviously one would like to generalize some
of these results to higher dimensions.

The notions of equivalence and direct sums extend in a
straightforward fashion to
dimensions higher than 2.  As we have seen any two-dimensional
unitary TQFT can be written as a direct sum of theories for which the
space associated with the circle is 1-dimensional.  It is appropriate to
call such theories irreducible.  It seems to be of importance to
introduce a notion of irreducibility and to establish
associated decomposition properties for
TQFT's in general.
A large class of 3-dimensional TQFT's has been constructed
\cite{durhuus,durhuus2,felder,chung}
generalizing the
theory of Turaev and Viro \cite{tv} and these theories
serve as candidates for irreducible theories since the dimension of
the Hilbert space of the sphere is always 1 and thus they cannot
be decomposed.

A natural question to ask is whether it is true in higher dimensions that
the partition functions for closed manifolds determine a theory up to
equivalence as is the case in two dimensions.  If the partition functions
for manifolds $M$ with $\pa M=\Sigma$ span the Hilbert space associated to
$\Sigma$ for all boundaries $\Sigma$, then it is not hard to show
that this is the case.  However, this condition is in general not
fulfilled.  In the two-dimensional case it holds only if the spectrum of the
handle operator is non-degenerate.

\end{document}